\documentclass[twocolumn,numberedappendix,iop]{emulateapj}
\shorttitle{The Herschel Cold Debris Disks}
\shortauthors{G\'asp\'ar \& Rieke}
\submitted{\today}
\journalinfo{The Astrophysical Journal}

\usepackage{apjfonts}
\usepackage{amsmath}
\usepackage{multirow}

\usepackage{color}
\usepackage{ulem}
\usepackage{cancel}

\newenvironment{packed_item}{
\begin{itemize}
  \setlength{\itemsep}{1pt}
  \setlength{\parskip}{0pt}
  \setlength{\parsep}{0pt}
}{\end{itemize}}
      
\begin{document}

\title{The Herschel Cold Debris Disks:\\Confusion with the Extragalactic Background at $160~\micron$}

\author{Andr\'as G\'asp\'ar}
\author{George H.~Rieke}
\affil{Steward Observatory, University of Arizona, Tucson, AZ, 85721}

\begin{abstract}
The Herschel ``DUst around NEarby Stars (DUNES)'' survey has found a number of debris disk candidates
that are apparently very cold, with temperatures near 22K. It has proven difficult to fit their spectral 
energy distributions with conventional models for debris disks. Given this issue we 
carefully examine the alternative explanation, that the detections arise from confusion 
with IR cirrus and/or background galaxies that are not physically associated with the 
foreground stars. We find that such an explanation is consistent with all of these detections. 
\end{abstract}

\section{Introduction}

Debris disks play a vital role in our understanding of the exterior parts of
planetary systems. While inner orbit ($< 5~{\rm AU}$) planets are now readily 
observed with various techniques (i.e., radial velocity and planetary transit
surveys), wider orbit planets are significantly more difficult to detect, with
only a handful of them discovered by direct imaging \citep{marois08,kalas08,lagrange10,rameau13,currie14,kraus14,bailey14}. 
However, with their large surface areas, even low mass and low density debris disks 
are relatively easy to detect in the mid- to far-infrared wavelengths at larger
stellocentric distances, providing ways to study the outer parts of the systems.

Debris disks have a number of characteristic temperatures, of which the most prominent are 
$190~{\rm K}$ \citep{morales11} and 45 -- $80~{\rm K}$, with a weak dependence on the spectral 
type of the star \citep[e.g.,][]{ballering13}. Our solar system is an example, with the
Asteroid belt at $~2.3 - 3.3~{\rm AU}$ and Kuiper belt at $30 - 50~{\rm AU}$ \citep{backman95,vitense12}.

An intriguing new result from {\it Herschel} was the discovery of a new class 
of cold debris disks \citep{eiroa11,eiroa13}, with characteristic temperatures
of $22~{\rm K}$. Scaling from the models for Kuiper Belt dust by \cite{yamamoto98}, 
such a disk would be located at about 120 AU, in an environment dramatically different 
from those normally assumed for debris disks. The properties of such disks have been 
studied by \cite{krivov13}, who concluded that they would need to be made up of particles
that are larger than a few millimeters and smaller than $10~{\rm km}$, 
that are dynamically quiescent, and have orbital eccentricities and inclinations $ \le 0.01$.
Systems with such specific parameters are not just difficult to form, but also challenging to maintain,
when one considers all the destructive external effects disks at large stellocentric regions (especially ones
outside the ``stello-pause'') may experience (e.g., erosion by the interstellar medium,
stellar fly-bys, etc.). 

In this paper, because of the issues detailed above, we re-evaluate the possibility 
that these excesses are not intrinsic to the stars but result from confusion with unrelated sources.
In section \ref{sec:false}, we show that both likely forms of confusion noise, IR cirrus and distant background galaxies, 
would match the apparent temperature of the cold excesses. In section \ref{sec:confusion}, we 
show that standard treatments of confusion noise suggest that such sources may significantly 
affect the apparent detection of cold debris disks. In section \ref{sec:MC}, we follow up this 
possibility with a Monte Carlo analysis, which we then use in section \ref{sec:stat} to evaluate 
the null hypothesis that the apparent cold disks are instead drawn from the populations of confusing sources. 
In section \ref{sec:params}, we investigate the dependence of the results on the interval of 
the background galaxy fluxes considered in the statistical analysis, while in section \ref{sec:comps},
we compare our results to previous statistical analyses. Finally, in section \ref{sec:sum}, we
summarize our results.

\section{Possible Sources of False Cold Disk Signatures}
\label{sec:false}

The cold debris disks have a characteristic temperature of $22~{\rm K}$ 
\citep{eiroa11}. At this temperature an excess by a factor of two at $160~\micron$
yields an excess by only a factor of 1.2 at $100~\micron$. That is, an excess below
typical detection limits at $100~\micron$ and shorter wavelengths can be substantially
above the stellar output at $160~\micron$. We now consider whether the SEDs of the 
possible confusing sources are consistent with this value. 

First, we consider confusion by infrared cirrus. There are a number of relevant measurements: 
1.) \cite{roy10} use BLAST data to find temperatures of $19.9\pm1.3~{\rm K}$ 
and $16.9\pm0.7~{\rm K}$ for cold interstellar dust in two 
regions; 2.) \cite{martin10} fit early Herschel data with a temperature 
of $23.6\pm1.0~{\rm K}$; 3.) \cite{bracco11} find $T = 19.0\pm2.4~{\rm K}$, 
using a different set of early Herschel data; 4.) 
\cite{veneziani13} use Bayesian methods with a broad set of data to find 
temperatures in the ISM cold dust of $\sim 20~{\rm K}$ with a range of 
about $4~{\rm K}$ around this value. Therefore, IR cirrus is a viable candidate 
to contribute to emission at the appropriate temperature for the apparent 
cold disks. 

We now turn to confusion by background galaxies. There are a number of systematic
changes in the SEDs of luminous galaxies with increasing redshift (and increasing
luminosity at the detection threshold \citep[e.g.,][and references therein]{rujopakarn13,berta13,symeonidis13}).
We have quantified these trends as in \cite{rujopakarn13}. We have fitted a blackbody 
to the appropriate galaxy SED for direct comparison with the assumed disk SED in 
\cite{eiroa11}. We take luminosities between the lower envelope of the distribution of
detection limits with redshift in \citeauthor{magnelli13} (\citeyear{magnelli13}, Figure 8) and twice this value,
to obtain luminosities as a function of redshift, characteristic of the faintest sources
detected with PACS. We then redshift blackbody fits to the SEDs for the appropriate
luminosities by the appropriate values to obtain apparent temperatures of faint 160 $\micron$
detections vs.\ redshift. We find that the values range from about 25 K at $z = 0.4$ to
about 29 K near $z = 0.8$, from where they decline to about 20 K at $z = 2$. \cite{magnelli13}
give redshifts of $z = 1.22^{+0.68}_{-0.41}$ and $z = 0.94^{+0.52}_{-0.38}$
(interquartile ranges) respectively for 160 $\micron$ sources fainter and brighter than
\mbox{2.5 mJy}. Thus, the faint detections should fall within the 20 - 29 K apparent
temperature range. The temperatures estimated from the 100 and 160 $\micron$
measurements of the six sources identified as having cold excesses in \cite{eiroa13}
range from 22.5 to 31 K for the three with probable weak 100 $\micron$ excesses
(HIP 73100, 92043, and 109378) and from 2 $\sigma$ upper limits of 21 to 26.5 K for
the three with no indicated 100 $\micron$ excesses (HIP 171, 29271, and 49908).
We conclude that the expected spectral behavior of faint background galaxies is 
consistent with the temperatures assigned to the cold debris disks.

\section{Estimates of the Effects of Confusion Noise}
\label{sec:confusion}

Confusion with distant background galaxies becomes an increasing issue with 
increasing wavelength for the Herschel instruments. A conventional 
definition of the confusion limit is the ``source density criterion (SDC)'',
when 10\% of the sources of a given flux are so tightly crowded that they 
cannot be measured. \cite{dole03} find that this limit corresponds to 
16.7 beams per source, where the definition of the beam area is based on 
that by \cite{condon74}. \cite{berta11} estimate that the SDC is reached
for Herschel at source flux densities of 0.4, 1.5-2, and $8~{\rm mJy}$ respectively 
at 70, 100, and $160~\micron$. Given the fall of a stellar photospheric output 
inversely as the square of the wavelength, the SDC is a significant issue for the \cite{eiroa13} sample
only at the longest wavelength band of these three. However, the flux 
densities attributed to the cold disks are very similar to the limit there. 
Of the $\sim 100$ sources without $100~\micron$ excesses, the confusion statistic 
would imply that roughly 6 would be confused with background galaxies at $8~{\rm mJy}$ or 
brighter, compared with the six sources
identified by \cite{eiroa13} as cold disk sources, some
of which have excess fluxes less than $8~{\rm mJy}$. This agreement calls for a more detailed
investigation.

\cite{hogg98} show that sources near the confusion limit and with 
low ratios of signal to noise tend to be biased too high in apparent 
brightness. They derive a correction dependent on the slope of the source 
counts and the signal to noise ratio of the source, to remove this bias and 
assign the maximum likelihood flux to a source. In the case of the $160~\micron$ galaxy
measurements, a slope of $q=0.9$ can be derived for the source counts from the 
\cite{magnelli13} data between 1 and 10 mJy, while the signal to noise ratios can be obtained from \cite{eiroa13}, 
Table 14. Table 1 shows the six cold disk candidate stars with the resulting estimates of the fluxes
from the disks alone at $160~\micron$. We have left the error estimates as in \cite{eiroa13}, although 
\cite{hogg98} argue that the errors should be expected to increase
in these nearly-confusion-limited cases \citep[see their Figure 2 and also Figure 
3 in][]{hogg01}. Four of the six candidates have dropped below the usual
$\chi_{160} > 3$ detection criterion\footnote{$\chi_{160} = \left(F_{160} - P_{160}\right)/\sigma_{160}$, 
where $F$ is the measured flux, $P$ is the estimated photosphere, and $\sigma$ is the
error of photometry.}, suggesting that a more 
detailed treatment of the confusion effects may be critical in evaluating 
the reality of the cold disks.

\section{Monte Carlo Analyses}
\label{sec:MC}

The preceding sections indicate that confusion noise may play a significant 
role in mimicking the signature of a hypothetical 
extremely cold debris disk. We therefore perform various Monte Carlo analyses, 
allowing a relatively easy exploration of the confusion noise in more detail within the
full parameter space. Our analyses considers two main sources of obtained flux: 
cirrus noise and background galaxies. We detail these in the following subsections, 
while in Table \ref{tab:var} we summarize the parameters of the analyses with the 
default values given. The numerical variables of the model (i.e. the size
of the artificial field, log bin size in the galaxy distribution, Airy pattern bin size)
were determined with convergence tests to ensure fast computational speeds with reliable 
results.

\begin{deluxetable}{llr}
\tablewidth{0pt}
\tablecolumns{3}
\tabletypesize{\scriptsize}
\tablecaption{Parameters of the Monte Carlo analysis\label{tab:var}.}
\tablehead{
\colhead{Variable} & \colhead{Description} & \colhead{Fiducial value}}
\startdata
D			& Size of artificial field			& $0.5~{\rm sq.\ deg.}$\\
$G_{\rm min}$		& Minimum galaxy flux considered		& $1~{\rm mJy}$\\
$G_{\rm max}$		& Maximum galaxy flux considered		& $225.42~{\rm mJy}$\\
$b_{\rm gal}$		& log bin size in the galaxy distribution	& $0.02$\\
$\sigma_{\rm cirrus}$	& Std.\ dev.\ of cirrus noise			& $0.505~{\rm mJy}$\\
$\mu_{\rm limit}$	& Location par.\ of noise log-norm distr.\	& 0.67 \\
$\sigma_{\rm limit}$	& Scale par.\ of noise log-norm distr.\		& 0.33 \\
$N_{\ast}$		& Number of positions tested			& $10^6$ \\
$r_t$			& Target radius					& $6\arcsec$ \\
$r_p$			& Photometry radius				& $8\arcsec$ \\
$S_{\rm in}$		& Sky aperture inner radius			& $18\arcsec$ \\
$S_{\rm out}$		& Sky aperture outer radius			& $28\arcsec$ \\
$\Delta B_{\ast}$	& Bin size in photometry distribution		& $0.1~{\rm mJy}$
\enddata
\end{deluxetable}

\subsection{Cirrus noise}

Determining the value of the 
cirrus noise is difficult. Because of this, our goal was to assign a highly conservative value to it, 
without neglecting it. This was also appropriate, as the DUNES survey was designed to observe sources in low cirrus 
background regions. We used Equation 22 of \cite{miville07} to calculate the confusion noise, which
is based on the power-spectrum of the far-infrared dust emission and calibrated to low levels.
The spectral index in the equation is given by Equation 4 in their paper. Using HSpot, we estimated the
average ISM flux background for the DUNES sources to be $<I_{160}> = 7.02~{\rm MJy}~{\rm sr}^{-1}$, and
a $<I_{160}>/<I_{100}>$ ratio of 1.845. Assuming the standard \cite{condon74} definition of beam size,
we derived a cirrus noise of $\sigma_{\rm cirrus} = 0.505~{\rm mJy}$.
This estimate is only about half as large as is indicated in the scaling relations in HSpot.
We dealt with the small number of sources with much stronger than average cirrus by eliminating them 
from our test sample, rather than trying to estimate the cirrus noise more accurately. In the Monte
Carlo simulations the cirrus noise value at each
test location was determined by choosing a value following a Gaussian probability function centered
at zero with a standard deviation of $\sigma_{\rm cirrus}$.

\begin{deluxetable*}{l|c|cc|cc}
\tablewidth{0pt}
\tablecolumns{6}
\tabletypesize{\scriptsize}
\tablecaption{The number of cold disk sources observed and predicted at various target radii. \label{tab:stats}}
\tablehead{
 & & \multicolumn{2}{c}{Data Realization} & \multicolumn{2}{c}{CDF} \\
 & \colhead{Observed} & \colhead{Point-source\tablenotemark{$\dagger$}} & \colhead{PSF Smoothed\tablenotemark{$\dagger$}} & \colhead{Point-source} & \colhead{PSF-smoothed}}
\startdata
$6^{\prime\prime}$ Target Radius	& 6 & 6.7 & 6.4 & 6.1 & 5.6 \\
$7^{\prime\prime}$ Target Radius	& 6 & 8.6 & 7.4 & 7.9 & 6.5 \\
$8^{\prime\prime}$ Target Radius	& 7 & 10.6   & 8.3 & 9.9 & 7.3
\enddata
\tablenotetext{$\dagger$}{Number of sources predicted at the location of the peak of the probability distribution in the data realization analysis (section \ref{sec:real}).}
\end{deluxetable*}

\subsection{Background galaxy contribution}

As introduced in section \ref{sec:confusion}, background galaxies can dominate
the confusion noise at far-IR wavelengths. For our Monte Carlo analyses, we randomly distributed
galaxies on a 0.5 sq.\ degree area, with a fiducial galaxy flux interval of 1 to $225~{\rm mJy}$,
although for certain tests we extended the lower limit to $0.012~{\rm mJy}$.
The galaxy number counts were adopted from three separate studies. Between 0.012 and $1.25~{\rm mJy}$
we used the modeling results found in Table B.2 of \cite{franceschini10}, between 1.42 and $28.38~{\rm mJy}$
we adopted the observed galaxy counts of the GOODS-S ultradeep Herschel survey from \cite{magnelli13}, while 
the number counts of the brightest galaxies were adopted from Table 5 of \cite{berta11} (all fields combined).
Although these are three independent studies, their differential number count curves connect 
smoothly. Total number counts were calculated in logarithmically evenly spaced flux bins, 
with the number counts appropriately interpolated (in log space) at the bin boundaries and integrated 
(also in log space) with a simple second order trapezoid method. Between 1 and $225~{\rm mJy}$ the 
artificial field (0.5 sq.\ degree) has altogether 19146 galaxies, and between 
6 and $13~{\rm mJy}$ it has $\sim 2776$ galaxies (or $\sim 5552$ galaxies per sq.\ degree), which agrees with the 
estimated 5500 galaxies per sq.\ degree in this interval cited by \cite{krivov13}. 

When considering confusion with background galaxies, \cite{eiroa13} only used the differential count value 
determined at $6~{\rm mJy}$, resulting in a smaller total number of estimated background sources 
($\sim 2000$ per sq.\ degree). As detailed in section \ref{sec:params}, one of the key differences between our analyses 
and the previous ones is that we use a larger interval of background galaxy fluxes (and integrate the 
differential distribution). As we will show, galaxies fainter than the detection threshold ($\le 6~{\rm mJy}$) 
contribute to the confusion noise, as their spatial distribution is not isotropic enough for their contribution 
to the total flux to be canceled out by sky subtraction, even when considering a completely random field as we do here. 
Natural clustering of galaxies will likely even enhance their contributions \citep{fernandez08}.

The results of our model will depend predominantly only on a single parameter, the beam solid angle ($\Omega$) 
(i.e., the confusion beam size). The value of the beam solid angle is a matter of definition. 
The classic \cite{condon74} definition of the effective beam solid angle is
\begin{equation}
\Omega_e = (\frac{1}{4}\pi\Theta_1\Theta_2)\frac{1}{\left(\gamma-1\right)\ln 2}\;,
\end{equation}
where $\Theta_1$ and $\Theta_2$ are the half-power axes of the elliptical Gaussian beam and $\gamma$ is
the slope of the differential distribution of sources. According to Table 3.1 of the PACS Observer's Manual,
$\Theta_1=10.65^{\prime\prime}$ and $\Theta_2=12.13^{\prime\prime}$ at a scan speed of 
$20^{\prime\prime}~{\rm s}^{-1}$, while the value of $\gamma$ is around
1.9 at low fluxes ($1$ - $10~{\rm mJy}$), according to the \cite{magnelli13} data. These yield an
effective beam solid angle of $162^{\prime\prime 2}$, or a confusion beam radius of $7.19^{\prime\prime}$.

The DUNES team uses the images to identify potentially confusing sources (of similar brightness to 
the target) at $\ge 6^{\prime\prime}$. Hereafter, we refer to this distance as the {\it target radius}. After excluding 
targets with confusing sources, they perform photometry in a photometry radius of $8^{\prime\prime}$. 
The sky background was subtracted based on a value measured in an annulus outside the photometry radius.
Our models were constructed to reproduce this measurement strategy. We assumed aperture photometry carried out within a 
radius of $8^{\prime\prime}$. We tested a variety of target radii inward of which we assumed it was no longer possible to distinguish 
a background source from the target, besides the $6^{\prime\prime}$ assumed by DUNES. In all models, we rejected 
targets with sources lying in the annulus between the target and photometry radii that also were more than 
$2.5 \sigma$ above a value chosen with a log-norm probability function with $\mu_{\rm limit}$ and 
$\sigma_{\rm limit}$ parameters (see Table \ref{tab:var}) that describes the distribution of photometry 
errors for the DUNES sample.

We integrate the flux of the sources within the photometry aperture 
and the corresponding sky annulus using two methods: treating the galaxies as point sources in one of them, 
and convolving their emissions with the Herschel Airy pattern at 160 $\micron$ in the other. The first is the 
traditionally used method when considering confusion, however, we have found that smoothing the emissions with 
the point spread functions (PSF) will affect the results of the confusion estimates. We introduce the results 
of both calculations for completeness and also to allow comparisons with previous work.

After generating the artificial background galaxy map, our code determined random
positions and performed the previously described ``aperture photometry''. 
For the smoothed model all partial fluxes contained within the apertures were included (i.e., fluxes
from sources both within and outside the apertures). For the point source
method the total fluxes of sources within the apertures were added to determine the total flux, but only for
sources that were located within the apertures.
To censor bright galaxies within the sky annuli, as was done by \cite{eiroa13}, we removed bright 
galaxies from the sky annulus above a simulated upper limit. As with the photometry aperture, this 
limit was set at $2.5\times$ a $\sigma$ value that was randomly chosen from the
photometry error distribution described above. Finally, the flux within the sky background was 
normalized by the ratio of the aperture area to the sky annulus area. 
The flux at the test location was then de\-ter\-min\-ed by subtracting the ``sky background'' from the flux
determined within the confusion beam aperture and adding the cirrus noise. 

\section{Statistics}
\label{sec:stat}

In this section, we compare the results of our model to the DUNES observations. 
We first define the DUNES sample we compare our models to, and then compare the 
model results to the observations with various statistical methods, while varying the target radius.
In Table \ref{tab:stats}, we summarize the detection statistics of the observations and the models.

\begin{figure*}
\begin{center}  
\includegraphics[angle=0,scale=0.68]{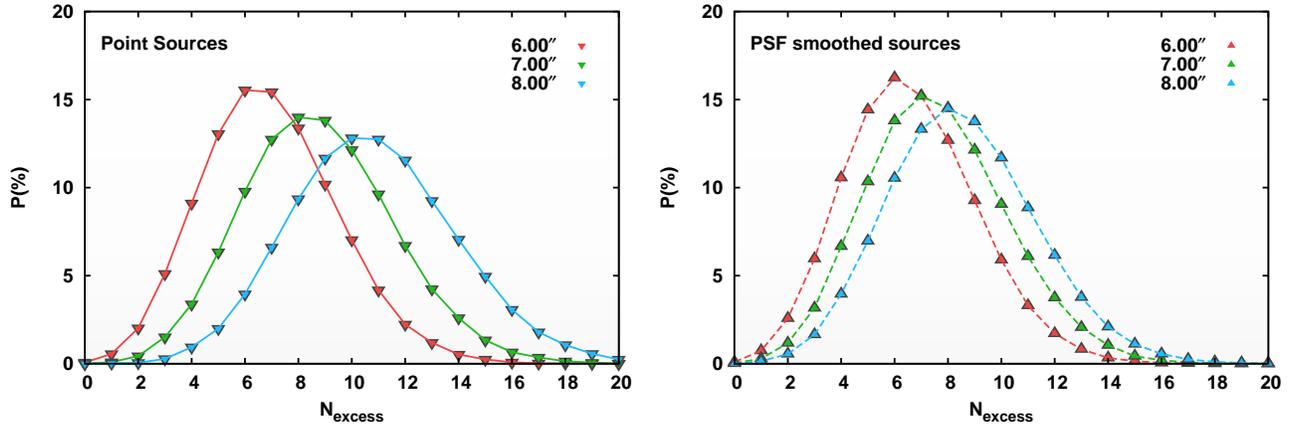}
\caption{The probabilities of detecting certain number of cold sources at various target radii, using the data realization method described in section \ref{sec:real}. In the {\it left panel},
we show the results of the ``point sources'' model and in the {\it right panel}, we show the results of the
``PSF smoothed sources'' model.}
\label{fig:sampling}
\end{center}
\end{figure*}

\subsection{The DUNES sample}
\label{sec:sample}

There are 133 sources in the DUNES sample \citep{eiroa13} of 
which 131 have data at $160~\micron$. Of these, 100 sources do not have detectable excesses at either PACS wavelengths.
From these 100, we removed 6 sources whose limits on their 
$160~\micron$ excess were higher than the typical value within the sample 
(HIP 71681, HIP 71683, HIP 88601, HIP 104214, HIP 104217, and HIP 108870).
Of the remaining 31 excess sources, \cite{eiroa13} list 6 as harboring cold debris disks.
Although it is listed as a cold disk candidate, the excess for HIP 92043 is detected at 
$70~\micron$ (both MIPS and PACS) and at $100~\micron$ and $160~\micron$ \citep{eiroa13}, 
so its identification as a cold disk candidate depends on the relatively weak $160~\micron$ result in Table 1. 
We computed a weighted average of the MIPS and PACS $70~\micron$ data, obtaining an excess of $11.9 \pm 3.3~{\rm mJy}$,
took the $100~\micron$ result from \cite{eiroa13} and the maximum likelihood value at
$160~\micron$ from Table \ref{tab:george} and then fitted the excess spectral energy distribution 
at all three wavelengths with a modified blackbody with $\beta = 0.65$ \citep{gaspar12b}. We found that a disk temperature of $62~{\rm K}$ fitted within the errors 
($\chi^2_{\rm reduced} = 1.35$), so there is no need to hypothesize a cold disk for this star and we remove 
it from the cold disk sample. Of the original 6 cold debris disk candidates \citep{eiroa13},
we only consider HIP 171, HIP 29271, and HIP 49908 most likely to have alternative explanations
for their apparent far infrared excesses. HIP 73100 and HIP 109378 show evidence for excess emission
at $100~\micron$, but the rapid increase in their SEDs to $160~\micron$ probably arises from confusion.

\begin{deluxetable}{cccc}
\tablecolumns{4}
\tablewidth{0pt}
\tablecaption{Maximum Likelihood $160~\micron$ Flux Density Estimates \label{tab:george}}
\tablehead{
\colhead{Star} & \colhead{Max. Likelihood $160~\micron$ disk flux density} & \colhead{Error} & \colhead{$\chi_{160}$}\\
\colhead{HIP} & \colhead{(mJy)} & \colhead{(mJy)} & \colhead{}
}
\startdata
171& 6.2& 2.5& 2.5 \\
29271& 6.0& 2.2& 2.7 \\
49908& 4.8& 2& 2.4 \\
73100& 8.3& 2.5& 3.3 \\
92043& 9.3& 4& 2.3 \\
109378& 9.8& 2& 4.9
\enddata
\end{deluxetable}

Apart from the five cold disk candidates, additional
spurious sources were listed in Table D.1 of \cite{eiroa13}. Two of the spurious sources have heavy cirrus
contamination [HIP 29568 (``structured background'') and HIP 71908 (``emission strip''; HSpot indicates a high 
interstellar background level of $59.7~{\rm MJy}~{\rm sr}^{-1}$)] and one is probably a spurious detection (HIP 38784).
Four of the remaining sources (HIP 40843, 85295, 105312, and 113576) are potentially contaminated by background 
galaxies. Of these four sources, two have the peaks of the emission of their $160~\micron$ component within the
$8^{\prime\prime}$ photometry aperture of the survey (HIP 85295 at $4.8^{\prime\prime}$ and HIP 105312 
at $7.16^{\prime\prime}$). For our models to stay consistent with the observational sample, we include these 
two sources in the cold disk sample (one or two of them, depending on the size of the target radius). 
This means that there are a total of six/seven sources with apparent cold excesses (HIP 171, HIP 29271, 
HIP 49908, HIP 73100, HIP 85295, HIP 105312, and HIP 109378).

This leaves us a total sample of 93/94 sources (6/7 with excess and 87 without), depending on the considered
target radius, with the boundary at $7.16^{\prime\prime}$. Of the non-excess
sources, 33 have measured fluxes, while the remaining 54 only have upper limits. The remaining 25 
sources with detected debris disk excesses were not included in the statistical analysis, as estimating a possible
level of contamination for them is not possible. Of these sources, three (HIP 4148, HIP 27887, and HIP 51502) 
have equilibrium temperatures around 30 K, close to the levels of the cold disk candidates. The final results of
the paper would indicate an additional 1.98 cold sources remaining in the sample of 25, possibly also explaining
the far-IR excesses observed at these three sources.

\begin{figure*}
\begin{center}
\includegraphics[angle=0,scale=0.68]{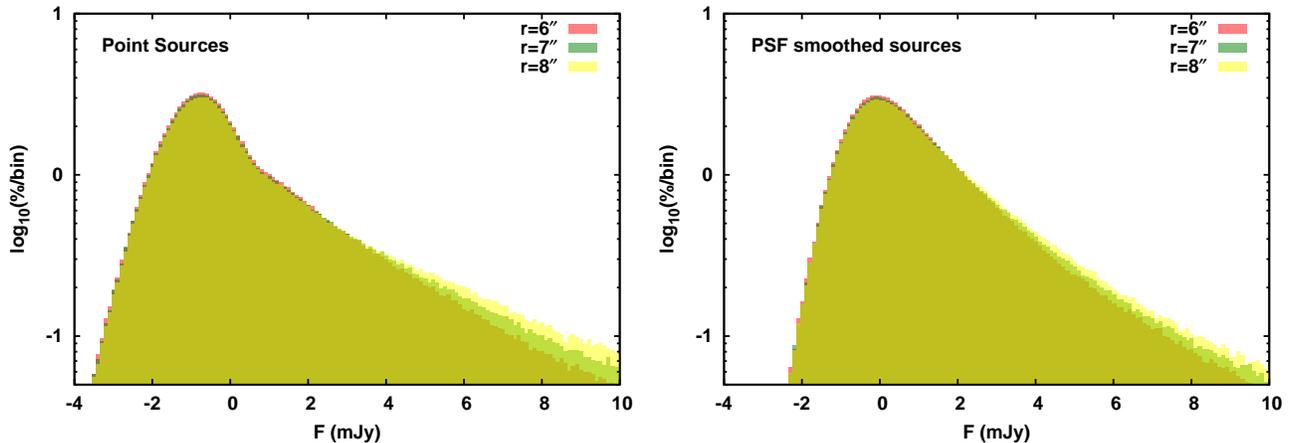}
\caption{The distributions of background fluxes assuming the fiducial values
of the model, varying only the confusion beam radius. In the {\it left panel},
we show the results of the ``point sources'' model and in the {\it right panel}, we show the results of the
``PSF smoothed sources'' model.}
\label{fig:DF}
\end{center}
\end{figure*}

\subsection{Method 1: Via Realization of Data}
\label{sec:real}

The first method we apply realizes artificial datasets and counts the number of detections within the
dataset. First, 93/94 source locations are randomly selected within our artificial field and the total 
flux at these locations calculated according to the procedure described in section \ref{sec:MC}. 
Then a detection threshold (determined at $3\sigma_F$) is randomly paired to each artificial 
location from the sample of 93/94 DUNES sources. If the total flux is larger than the detection threshold
then the number of detections in the realized dataset is incremented by one. We realized $10^5$
datasets of 93/94 sources for each of the tested target radii. In Figure \ref{fig:sampling},
we show the results of these tests for both the ``point sources'' and ``PSF
smoothed sources'' models and in Table \ref{tab:stats}, we summarize the detection statistics of the model. 
The probability of finding more sources increases with larger target radii, as expected. 

The ``point-sources model'' yields probability curves that are strongly dependent on the target radii.
With over 40\% of the photometry area located between 6 and 8$^{\prime\prime}$, sensoring confusing
sources in the outer aperture is critical. The probability curves are wide, for example at a target
radius of $7^{\prime\prime}$, the model predicts $8.55\pm2.79$ sources, meaning that detecting 5.7 sources
is just as likely as detecting 11.3. 

A closer representation of the measurements is performed by the ``PSF smoothed-sources model''. The
distributions are narrower and the peaks are closer and at lower values than for the simpler ``point-sources
model''. The peaks shifting to lower values is due to the generally higher sky background values, which
is a result of contributions to the sky flux from sources outside the reference sky annulus, which
are now smoothed into the sky area. As the sky annulus is larger in area than the aperture photometry 
area, and it also receives contributions from sources inside of it as well as from outside of it, this
is a significant effect. Moreover, the distributions are also narrower due the PSF smoothing, as 
background levels become more homogenous. As an example, at a target radius of $7^{\prime\prime}$, the ``PSF smoothed-sources''
model predicts $7.38\pm2.61$ sources. This agrees well with the 6/7 sources expected at $7.16^{\prime\prime}$
according to the observations.

\begin{figure*}
\begin{center}
\includegraphics[angle=0,scale=0.68]{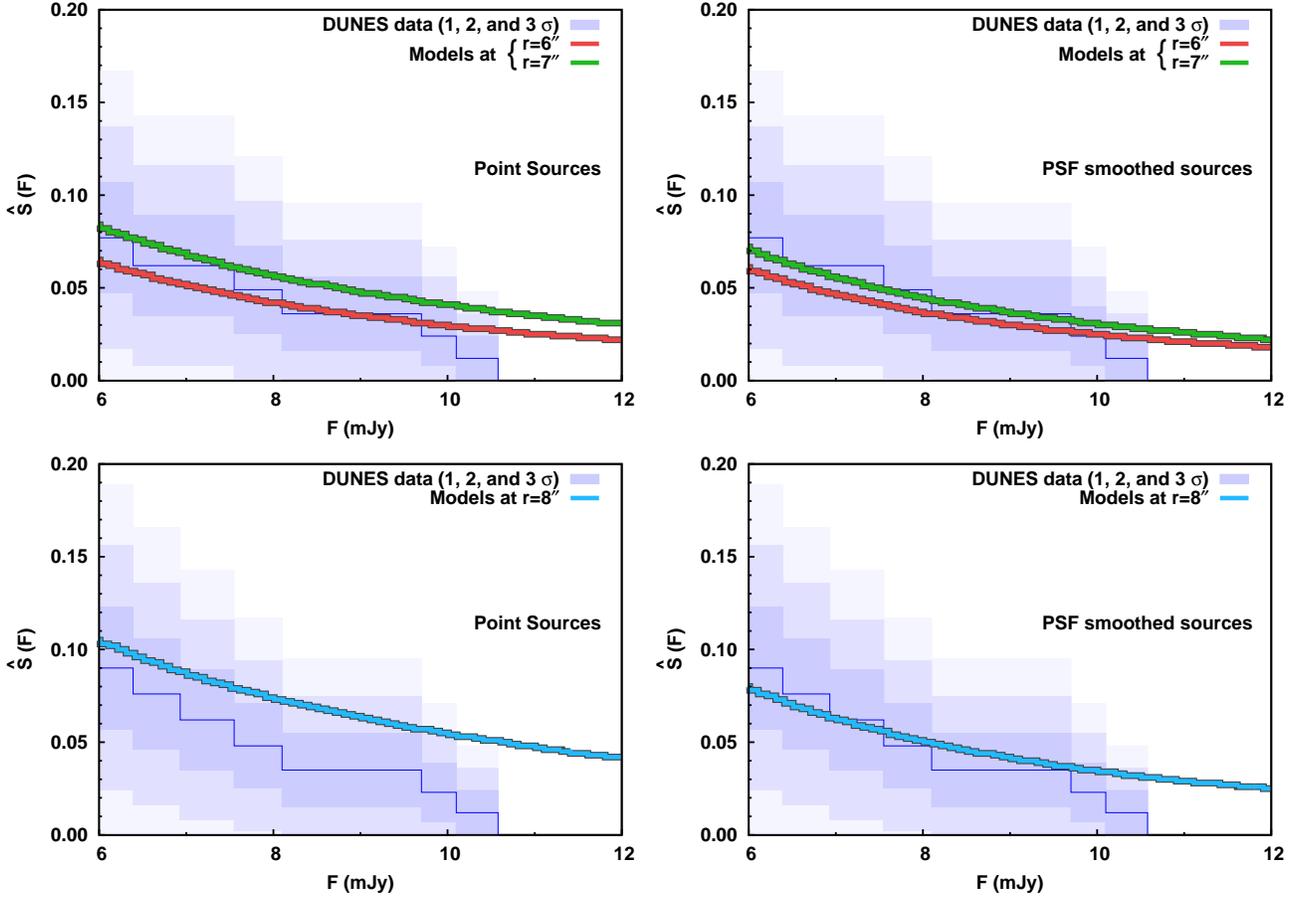}
\caption{The KM estimates (see section \ref{sec:KM}) of the observation and the models. The blue histograms
show the 1, 2, and 3 $\sigma$ confidences of the KM estimates for the observations. The $8^{\prime\prime}$
models ({\it bottom panels}) are compared to the 94 source DUNES sample, while the 6 and $7^{\prime\prime}$
models ({\it top panels}) are compared to the 93 source DUNES sample.}
\label{fig:SF}
\end{center}
\end{figure*}

The significant number of potentially confusing sources among the stars without excesses or with cold
ones raises a question of the contamination among those with debris disks. A rough estimate can be obtained
by noting that among the stars with normal disks, there are five with apparent detections at 160 $\micron$ 
($\chi_{160} > 3$) and with flux densities less than \mbox{20 mJy}, within the range where confusion is a
risk. From the statistics above, these numbers suggest that no more than one of the normal disk stars may
have a 160 $\micron$ flux density dominated by a background galaxy.

\subsection{Method 2: Via Distribution Functions}

With the second method, we generated distributions of the artificial fluxes by 
testing $N_{\ast}$ number of random positions. The flux values of our sample of $N_{\ast}$ 
test points were then binned with a bin size of $\Delta B_{\ast} = 0.1~{\rm mJy}$.
These distributions were then compared to the observed distribution of fluxes
with various methods.

\subsubsection{Percentages with Cumulative Distributions}

The simplest test that can be performed is determining the percent of sources
above given thresholds using the cumulative distribution function (CDF) of the model \citep[as in][]{krivov13,eiroa13}.
This test is not rigorous (e.g., it does not take account of upper limits above the
sample detection threshold). However,
for illustration and to allow comparison with previous statistical analysis, 
we begin with the results of this test. In Figure \ref{fig:DF},
we show binned distribution functions of background fluxes of our nominal 
model, while varying the target radius, for both the ``point sources'' and ``PSF smoothed
sources'' models. Increasing the target radius, as expected,
will widen the distribution and yield more high flux sources. The faintest 
cold disk candidate has an excess of $6.39~{\rm mJy}$. The number of predicted sources
above this limit at various target radii is also summarized in Table \ref{tab:stats}.
The results of the CDF analysis compare fairly well to the observed number of sources with 
cold disk signatures, especially when considering the classic \cite{condon74} definition of 
confusion beam size and the more realistic ``PSF smoothed sources'' model. 

The peak of the distribution at negative values in Figure \ref{fig:DF} results because 
more brighter galaxies will be located within the larger area sky annulus than within the search area. 
Unless the area of the sky annulus is equal to the photometry aperture area, this will
always result in a negative bias. We have tested this by using a sky annulus with the
same area as the photometry aperture, resulting in a peak at zero. The effect is
less prominent in the ``PSF smoothed sources'' model compared with the ``point sources model''.

\subsubsection{Kaplan-Meier estimates}
\label{sec:KM}

The Kaplan-Meier (KM) estimates \citep{km58} of both the modeled and observed distributions
provide a systematic method to compare these distributions while taking account of the upper
limits (or censoring) in the observations. The KM method has been adopted for astronomical data 
analysis \citep[e.g.,][]{feigelson85}, where it is
useful for randomly picked datasets, such as the background distribution in the DUNES survey. We used
the ASURV package \citep{feigelson85} to calculate the KM estimates and compare the KM curves of the observations
and models at various target radii in Figure \ref{fig:SF}. The DUNES data we compare our models to
depends on the target radius with the addition of the extra seventh source when comparing to the $8^{\prime\prime}$
model. The bottom panels in the Figure show these calculations, while the top panels show the comparisons at 
$6$ and $7^{\prime\prime}$ with the KM curve of the observations using six excess sources.
For the observations, we have set all sources apart from the cold disk candidates (the remaining 87 sources) 
as upper limits. The upper limits were
set to ${\rm UL} = F-P+3\sigma_F$ for sources where the photospheres were detected and kept at their original
published upper limit value minus the estimated photosphere where they were not. Here, $F$ is the measured
flux density, $P$ is the expected value from the stellar photosphere, and $\sigma_F$ is the quoted uncertainty.
The models generally appear to agree closely with the distribution of the observations.

\subsubsection{Kolmogorov-Smirnov test on the incompleteness-corrected sample}
\label{sec:KSKM}

\begin{figure}
\begin{center}
\includegraphics[angle=0,scale=0.68]{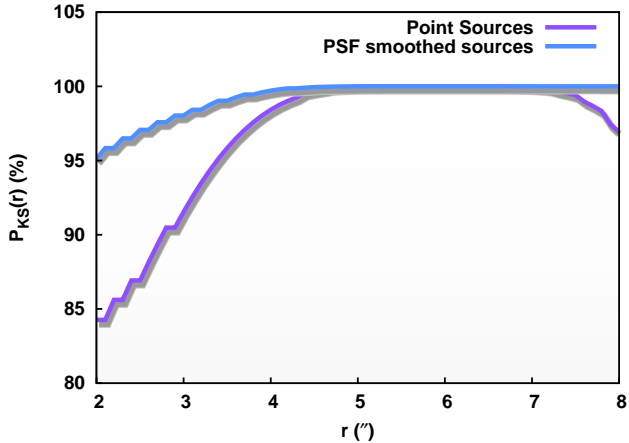}
\caption{Probability of agreement between the incompleteness corrected data 
and the model as a function of the confusion beam radius.}
\label{fig:KSKM}
\end{center}
\end{figure}

There is no standard method to determine the probability of agreement between censored data and a numerical 
model. We have therefore proceeded as follows. The Kaplan-Meier estimator introduced in the previous subsection
can be thought of as an incompleteness-corrected CDF, as it carries on the probabilities 
of previous events occurring with the knowledge of the censoring. To obtain a rigorous test making use of the 
upper limits, we perform a Kolmogorov-Smirnov (KS) test on the incompleteness-corrected sample,
by increasing the weight of the surviving sample members exactly the way an incompleteness correction would.
The KS statistic was only calculated for sources above the
detection threshold of $6.39~{\rm mJy}$ (as below we do not have any data) and the probabilities were 
calculated by scaling with the complete distribution. In Figure \ref{fig:KSKM}, we show the
probabilities obtained as a function of the target radius with these methods. The probability 
curve indicates that the data are consistent with being drawn from the confusion-limited model at 
$> 80\%$ confidence for all target radii between 2 and $8^{\prime\prime}$ (the drop in probability for 
large radii is because the model predicts too many detections, 
so this case is not of interest). This range of target radii includes all plausible definitions for the PACS beam. 
The figure shows that even when considering a smaller target radius, as long as the photometry is 
performed up to $8^{\prime\prime}$, the model results will be consistent with the observational statistics.

We also performed the Anderson-Darling K-sample test \citep{scholz87} on the incompleteness corrected 
sample, as it is more sensitive at the edges of the distributions than the KS test, using the statistical 
analysis software package {\tt R}. 
The observed data has many upper limits above the detection threshold
of $6.39~{\rm mJy}$, and hence has significant corrections for incompleteness. These 
corrections introduce pseudo-ties in the data, to which the Anderson-Darling test is
sensitive. Therefore, we used the method that assumes ties within the data, described in section 5 of their
paper. The analysis showed that the two distributions are indistinguishable within
target radii of $4.2$ and $6.3^{\prime\prime}$ for the ``point sources'' model and for all target radii larger than 
$5.3^{\prime\prime}$ for the ``PSF smoothed sources'' model at a 95\% confidence level. 

\section{Parameter dependence} 
\label{sec:params}

Although the results mainly depend on the choice of target radius, here we investigate how
the results depend on the range of galaxy fluxes considered. The main motivations for this study
are the previous analyses \citep{eiroa13,krivov13} that rejected the hypothesis that all of these systems 
could be explained by confusion, but only used a limited range of galaxy fluxes, between 6 and $13~{\rm mJy}$.

\begin{figure}
\begin{center}
\includegraphics[angle=0,scale=0.68]{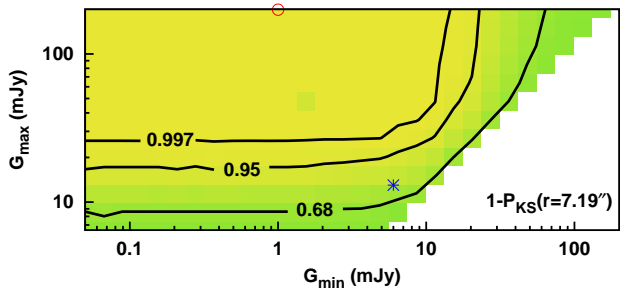}
\caption{The probability of agreement between the incompleteness corrected data and
the model as a function of the minimum and maximum galaxy fluxes considered in the model.
The blue star shows the interval considered by \cite{eiroa13} and \cite{krivov13} and the
red circle shows the interval considered by our nominal model.}
\label{fig:SminSmax}
\end{center}
\end{figure}

We simulated 900 models, with both the minimum and maximum galaxy fluxes ranging between 
0.012 and $225.49~{\rm mJy}$ and a target radius of $7.19^{\prime\prime}$ using the PSF smoothed
approach, and calculated the completeness corrected KS test (as in section \ref{sec:KSKM})
for each of them. In Figure \ref{fig:SminSmax}, we show the results of these KS tests as
a 2D plot, contouring the 1, 2, and 3 $\sigma$ probabilities and also plotting the ranges considered
by the previous studies and ours. Compared with the full-range estimate, the limited flux interval 
produces less sources through the omission of noise due to the cumulative effects of faint sources.
This result demonstrates that the difference between our work and the previous conclusions
about the cold disks can largely be explained by the inappropriate limitation in confusing
source fluxes assumed by \cite{eiroa13} and \cite{krivov13}.

\section{Comparison to previous work}
\label{sec:comps}

The DUNES and DEBRIS Herschel Open Time Key Program surveys were the first surveys ever
conducted with the specific goals of detecting debris disks at wavelengths between $100$ and $800~\micron$.
At these wavelengths, as shown in Section \ref{sec:false}, confusion with the extragalactic
background and/or infrared cirrus can be an important effect. While simple galactic number
count statistics suffices for confusion studies at shorter wavelengths, due to the larger
confusion beam sizes and the high number of confusing sources at the detection threshold,
a more sophisticated analysis is necessary at these longer wavelengths.

In the discovery paper, \cite{eiroa13} analyze the likelihood of these sources originating
from confusion with the extragalactic background in their Section 7.2.1. After excluding 
spurious sources with obvious high background/cirrus contamination, they conclude with a list
of six sources requiring an alternate explanation. Based on the \cite{berta11} galaxy counts
at $160~\micron$, they perform count statistics. Based on their artificial data
tests, they assume a confusion beam radius of $5^{\prime\prime}$, where they were able to separate
two equal sources with fluxes near the detection threshold value. For multiple sources that are
fainter than the detection threshold, this may be an inadequate confusion beam radius value,
especially when considering the classic \cite{condon74} definition of confusion beam size.
They also use the differential number density of galaxies at the detection threshold as a 
total  source count, yielding a low number of possible contaminating sources. In Section
\ref{sec:params}, we show the importance of using the full range of background galaxy fluxes
when calculating the effects of confusion. Finally, they considered their complete observational
catalog for the statistics (133 sources), including systems that were shown to harbor debris disks. 
Although systems with debris disks may also have background confusion at $160~\micron$, the contribution
from the background will be difficult to distinguish from the debris disk component, requiring these
systems to be removed from the analysis sample. These approximations resulted in a prediction
of only 1.2\% of the sources having background confusion. 

The theoretical analysis in \cite{krivov13} focused on explaining the physical likelihood
of cold debris disks existing and deem it unlikely that all of these sources could
originate from confusion with cirrus, which we agree with. They performed searches 
for strong X-ray and/or optical galactic counterparts, but the results from these tests
were largely inconclusive within the confusion beam. 
As in \cite{eiroa13}, they also performed statistical tests, mostly with the same arguments.
They show that the offsets between the assumed position of the sources and the $160~\micron$
fluxes are all within $5^{\prime\prime}$, however, as per the definition of confusion
beam ($7.19^{\prime\prime}$), all positions within it are not separable. This is also
noted in \cite{krivov13}, which is why they search for sources of background confusion 
within a radius of $6^{\prime\prime}$ in their statistical analysis. However, they only
look at extragalactic sources within the flux range of the cold sources (6 to 13 mJy),
not accounting for possible confusion originating from multiple fainter sources. They calculate
a confusion probability of 4.8\%, and scaling to the complete DUNES sample (133 sources)
predict 6.4 false detections. Assuming that all of the seven spurious sources in Table D.1
of \cite{eiroa13} are a result of extragalactic background contamination, they determine
that there is a 69\% probability that the remaining six cold disks are true 
debris detections. This argument, however, does not take into account that two of the
seven sources are obviously contaminated by high cirrus noise (HIP 29568 and HIP 71908),
while three of the remaining five (HIP 40843, HIP 105312, and HIP 113576) have the peaks
of their $160~\micron$ emission outside of the $6^{\prime\prime}$ confusion beam radius 
used in their analysis. Of the remaining two sources, HIP 38784 is a spurious detection
with double $160~\micron$ peaks (of which one is also outside of the $6^{\prime\prime}$
radius. There is only a single source from their Table D.1, HIP 85295, that needs to
be counted as a source in the statistical analysis, as in our paper. 

The most detailed work on confusion estimates for debris disk studies at longer
wavelengths were performed by \cite{sibthorpe13}. As a first step, they convolve
the \cite{berta11} number counts with Gaussians with various error estimates as
a way of accounting for the Eddington bias. They present two calculations,
one that calculates the probability of a single source producing the confusion and
one that calculates the probability of one or more sources producing it. They 
also introduce a Monte Carlo style algorithm to calculate the probability of confusion.
However, their algorithm considers the survey limiting flux density not just as
a detection threshold, but also as the minimum galaxy flux in the model. For
the $7^{\prime\prime}$ model, at $S_{\rm lim}=6.39~{\rm mJy}$, they predict a probability
of 7.8\%, which is close to the value given by our ``point-sources'' model. However,
the methods of sky subtraction are not introduced in the paper, therefore we
are unable to access the final results presented in it.

The cold disk candidate HIP 92043 is analyzed in \cite{marshall13}. They also include 
a statistical argument whether the source can plausibly be a cold disk source in their
section 4.1. Although they add a cold component to their model to fit at $160~\micron$,
they describe their excess detection at this wavelength as marginal.
Their statistical analysis is along the lines of that of \cite{eiroa13} and predict 1\% 
of the sources having a contamination at the 12.9 mJy level. As a comparison, our CDF
model predicts 2.1\% of the sources having a contamination above the 12.9 mJy level
for the ``point-sources'', and 1.8\% of them for the ``PSF smoothed-sources'' 
model at their assumed $11.3^{\prime\prime}$ target aperture. Our higher values are
due to the same effects as previously. They also cite the work of \cite{sibthorpe13},
however, only consider their model where confusion with a single bright source is 
calculated. The MC models of \cite{sibthorpe13} show higher probabilities of 
confusion than their single source confusion analytic estimates.

\section{Summary}
\label{sec:sum}

In this paper, we evaluate the hypothesis of a newly discovered class of 
{\it Herschel} cold debris disks \citep{eiroa11,eiroa13,krivov13}. We test whether 
the apparent temperature and flux distributions are instead consistent with confusion 
noise. Although this scenario has been considered by previous work, there are a few 
differences between our analyses:
\begin{packed_item}
\item we simulate confusion noise using the full relevant range of background galaxy fluxes and allow
for confusion from multiple sources,
\item we account for the smoothing of the emissions by the PSF of the telescope,
\item we develop an analysis method that accounts for the censorship of the data due to the 
limitations in signal to noise ratio in the DUNES $160~\micron$ data.
\end{packed_item}

We test the hypothesis that the distribution of cold debris disks is entirely 
due to confusion with background galaxies (after rejecting cases with elevated noise from IR cirrus). 
We evaluate the hypothesis as a function of the target radius used to measure sources at 
$160~\micron$. We find that there is a greater-than-80\% probability that the two distributions 
(confusion noise and the proposed cold debris disks) are indistinguishable, so long as the 
confision beam is between 2 and $8^{\prime\prime}$ in radius. This range of beam size 
includes all plausible values for the DUNES measurements. We conclude that the background confusion 
hypothesis is a viable alternative to the cold debris disk explanation for the $160~\micron$ 
detections of these sources.

\acknowledgments

We thank Benjamin Weiner, Bran\-don Kel\-ly, E\-ric Fe\-i\-gel\-son, and E\-wan Ca\-me\-ron for 
inputs on the statistical analysis as well as the Astrostatistics and Astroinformatics Portal (ASAIP) 
hosted by Pennsylvania State University. Support for this work was provided by NASA through Contract 
Number 1255094 issued by JPL/Caltech.

\end{document}